\documentclass{ws-procs9x6}

\newcommand{\be}{\begin{equation}}
\newcommand{\ee}{\end{equation}}
\newcommand{\bea}{\begin{eqnarray}}
\newcommand{\eea}{\end{eqnarray}}
\newcommand{\non}{\nonumber}
\newcommand{\ra}{\rangle}

\newcommand{\al}{\alpha}

\begin{document}

\title{Generalized Drinfeld polynomials for highest weight vectors 
of the Borel subalgebra of the $sl_2$ loop algebra}

\author{Tetsuo Deguchi}

\address{
Department of Physics, Ochanomizu University, \\
2-1-1 Ohtsuka, Bunkyo-ku\\ 
Tokyo, 112-8610, Japan\\ 
E-mail: deguchi@phys.ocha.ac.jp
}

\maketitle

\abstracts{
In a Borel subalgebra $U(B)$ of the $sl_2$ loop algebra,  
  we introduce a highest weight vector $\Psi$.  
We call such a  representation of $U(B)$ that is generated by $\Psi$ 
{\it highest weight}. 
We define a generalization of the Drinfeld polynomial for 
a finite-dimensional highest weight representation of $U(B)$.   
We show that every finite-dimensional highest weight representation 
of the Borel subalgebra is irreducible 
if the evaluation parameters are distinct.  
We also discuss the necessary and sufficient conditions 
for a finite-dimensional  highest weight representation 
of $U(B)$ to be irreducible.   }

\setcounter{section}{0} 
 \setcounter{equation}{0} 
 \renewcommand{\theequation}{1.\arabic{equation}}
\section{Introduction}

In the classical analogue of the Drinfeld 
realization of the quantum $sl_2$ loop algebra, 
$U_q(L(sl_2))$,  
the Drinfeld generators, ${\bar x}_k^{\pm}$ and 
${\bar h}_k$ for $k \in {\bf Z}$, 
 satisfy the following defining relations\cite{Chari-P1,Chari-P2,Drinfeld}:
\bea 
& & [\bar{h}_j, \bar{x}_{k}^{\pm} ] = \pm 2 \bar{x}_{j+k}^{\pm} \, , \quad 
[\bar{x}_j^{+}, \bar{x}_k^{-} ] = \bar{h}_{j+k} \, , 
\non \\
& & [{\bar h}_j, {\bar h}_{k} ]=0 \, , \quad  
[{\bar x}_j^{\pm}, {\bar x}_k^{\pm}] =0 \, , \quad  
{\rm for} \, \,  j, k \in {\bf Z}.
\label{CDR}
\eea
In a representation of $U(L(sl_2))$, a vector $\Omega$ is called 
{\it a highest weight vector} if  $\Omega$ 
is annihilated by generators ${\bar x}_{k}^{+}$ 
for all integers $k$ and such that 
$\Omega$ is a simultaneous eigenvector of every generator 
of the Cartan subalgebra, ${\bar h}_k$ ($k\in {\bf Z}$)   
\cite{Chari-P1,Chari-P2}.  
We call a representation of $U(L(sl_2))$ {\it highest weight} 
if it is generated by a highest weight vector. 
 For a  finite-dimensional irreducible representation   
we associate a unique polynomial 
through the highest weight ${\bar d}_k^{\pm}$. 
It is shown that 
any given irreducible highest weight representation is finite-dimensional 
if and only if it has the Drinfeld polynomial \cite{Chari-P1}.

Recently it was shown that the XXZ spin chain at roots of unity 
has the $sl_2$ loop algebra symmetry 
\cite{Regular,DFM,FM1,FM2,Odyssey}.
 Fabricius and McCoy has conjectured \cite{Odyssey} that 
every Bethe ansatz eigenstate should be highest weight of the $sl_2$ 
loop algebra, and also that the Drinfeld polynomial 
can be derived from the Bethe state. It is explicitly shown that 
regular XXZ Bethe states in some sectors are indeed highest weight 
\cite{Regular}. 
However, it is still nontrivial how to connect the highest weight vector 
with the Drinfeld polynomial. In fact, the Drinfeld polynomial is defined 
for an irreducible representation not for 
a highest weight vector \cite{Chari-P1}. Furthermore, 
there exist 
finite-dimensional highest weight representations 
that are reducible and indecomposable. 
It has  been  shown that a given highest weight representation 
is irreducible if the evaluation parameters are distinct 
\cite{Chari-P3,RAQIS05}.  Here, we shall define 
evaluation parameters in \S 3.  
Thanks to the theorem,  
we solve the connection problem at least for the case of 
distinct evaluation parameters.

In this paper, we discuss a generalization of the theorem to the case of 
a highest weight representation of a Borel subalgebra of $U(L(sl_2))$. 
The generalization should play a key role  
in the study of the spectral degeneracy of the XXZ spin chain 
under twisted boundary conditions \cite{twisted,Korff-twisted,Deguchi-Kudo}. 
Let us consider the subalgebra generated by generators  
${ h}_0$, ${ x}_0^{+}$ and ${ x}_{1}^{-}$ satisfying 
the relations (\ref{CDR}).  
We call it a Borel subalgebra of $U(L(sl_2))$, and denote it by 
${U(B)}$.  It has the following generators:    
\be 
{ h}_k \, , \, { x}_k^{+} \,  \,  
{\rm for} \, k \in {\bf Z}_{\ge 0} \, , 
\quad   
{ x}^{-}_k  \, \,  {\rm for} \,  k \in {\bf Z}_{> 0}.    
\ee
We define a highest weight vector 
of the Borel subalgebra $U(B)$ by such a vector $\Psi$ 
that satisfies the following relations:  
\be 
{ x}_k^{+} \Psi =  0 \, , \quad  
{ h}_{k} \Psi  =  { d}_k \Psi \, ,  
\quad {\rm for} \, \, k \in {\bf Z}_{\ge 0} \, . 
\ee 
We call the representation of $U(B)$
 generated by $\Psi$  {\it highest weight} and the set 
 $\{{ d}_k \}$ 
 the highest weight. 
Here we note that $d_0$ is not necessarily an integer, since 
$x_{-1}^{+}$ does not exist in $U(B)$. 
  In \S 2 of the present paper, 
we derive a useful recursive relation 
of ${ x}_k^{-}\Psi$  for $k \in {\bf Z}_{>0}$.  
In \S 3 we introduce  
 a generalization of the Drinfeld polynomial 
for a finite-dimensional highest weight representation 
of the Borel subalgebra $U(B)$. 
In \S 4 we show 
that every highest weight representation of the Borel subalgebra 
with distinct and nonzero evaluation parameters 
is irreducible.

Throughout the paper, we denote by $\Psi$ 
 a highest weight vector of the Borel subalgebra $U(B)$ with 
 highest weight ${ d}_k$ 
 and by $V_B$ the representation generated by it, i.e. 
 $V_B= U(B) \Psi$. We also assume that $V_B$ is finite-dimensional. 

\setcounter{section}{1} 
 \setcounter{equation}{0} 
 \renewcommand{\theequation}{2.\arabic{equation}}
\section{Sectors of $V_B$ and nilpotency} 

\begin{lemma}
 Let us define the sector of ${h}_0=d_0 -2n$ 
 in $V_B$ for an integer $n \ge 0$ 
by the subspace consisting of vectors $v_n \in V_B$  
such that ${ h}_0 \, v_n= (d_0 - 2n) \, v_n$. 
Here we recall  ${h}_0 \Psi = d_0 \Psi$.   
 Then, $V_B$ is given by  
the direct sum of such sectors.  
Any vector $v_n$ in the sector of ${ h}_0 = d_0 -2n$ is expressed as  
a linear combination of monomial vectors 
${ x}_{j_1}^{-} \cdots { x}_{j_n}^{-} \, \Psi$.  
\label{lem:sector} 
\end{lemma} 
\begin{proof}    
It is clear from the PBW theorem \cite{Jacobson}. 
\end{proof} 

We note that generator ${x}_1^{-}$ is nilpotent in any $V_B$. 

\begin{definition}
We say that generator ${ x}_1^{-}$ is 
nilpotent of degree $r$ in $V_B$,  
if $({ x}_1^{-})^{r+1}  \Psi  = 0$,  
while $({ x}_1^{-})^j \Psi \ne 0$  for $0 < j \le r$.  
\end{definition}

The degree $r$ of nilpotency for generator ${ x}_1^{-}$  
gives the largest $n$ for non-vanishing sectors of $h_0=d_0 -2n$, 
as shown in the next proposition. 

\begin{proposition}  
If generator ${ x}_1^{-}$ is nilpotent of degree $r$,  
then the sector of ${ h}=d_0-2r$ is one-dimensional: 
every monomial vector in the sector is proportional to 
$({ x}_1^{-})^r  \Psi$ 
with some constant $C_{k_1, \ldots, k_r}$: 
\be 
{ x}_{k_1}^{-} \cdots { x}_{k_r}^{-}  \Psi  = 
  C_{k_1, \ldots, k_r} \, ({ x}_1^{-})^r  \Psi \,  , 
\, \,  {\rm for} \, \,  k_1, \ldots, k_r \in {\bm Z}_{> 0} \, .  
\label{eq:1-dim}
\ee
Furthermore,  sectors of ${h}=d_0-2n$ for $n > r$ are of 
zero-dimensional. For instance, we have  
${ x}_{k_1}^{-} \cdots { x}_{k_{r+1}}^{-}  \Psi  = 0$ for 
$k_1, \cdots, k_{r+1} \in {\bm Z}_{> 0}$ . 
 \label{prop:nilpotent}
\end{proposition}
\begin{proof}
Setting $m=r$  in lemma \ref{lem:inductive},  
we have eq. (\ref{eq:1-dim}).   
For the case of $n > r$ we show it from lemma \ref{lem:inductive} 
where we set $m=n$. 
\end{proof}

Let  $B_{+}$ be such a subalgebra of $U(B)$ that is  
generated by ${ x}_k^{+}$ for $k \in {\bf Z}_{>0}$.  
We define $(X)^{(n)}$ by $X^{n}=X^n/n!$. 
\begin{lemma}
Let $m$ and $t$ be integers satisfying $0 \le t \le  m+1$.  
In the Borel subalgebra $U(B)$, 
for  $k_1, \ldots, k_t, n \in {\bm Z}_{>0}$, and 
$\ell \in {\bm Z}_{\ge 0}$, 
we have 
\bea 
& & { x}_{\ell}^{+} ({ x}_{n}^{-})^{(m+1-t)} 
{ x}_{k_1}^{-} \cdots { x}_{k_t}^{-} \non \\
& = & - { x}_{\ell + 2n}^{-} ({ x}_n^{-})^{(m-t-1)} 
{ x}_{k_1}^{-} \cdots { x}_{k_t}^{-}
+ ({ x}_n^{-})^{(m-t)} 
{ x}_{k_1}^{-} \cdots { x}_{k_t}^{-} \,  
{ h}_{\ell + n}
\non \\
& & 
+ \sum_{j=1}^{t}  ({ x}_n^{-})^{(m+1-t)} 
\prod_{i=1, i \ne j}^{t}  { x}_{k_i}^{-} \, \cdot \,  
{ h}_{\ell + k_j}
+ (-2) \sum_{j=1}^{t} ({ x}_n^{-})^{(m-t)} 
{ x}_{\ell+n+ k_j}^{-} \prod_{i=1; i \ne j}^{t}  { x}_{k_i}^{-}  
\non \\
& &  + (-2) \sum_{1 \le j_1 < j_2 \le t} 
({ x}_n^{-})^{(m+1-t)} { x}_{\ell+ k_{j_1} + k_{j_2}}^{-} 
\prod_{i=1; i \ne j_1, j_2}^{t}  { x}_{k_i}^{-}  
\quad {\rm mod} \, U(B) B_{+}  
\label{eq:null-rel}
\eea 
\label{lem:null-rel} 
\end{lemma}

\begin{lemma} 
Suppose that ${ x}_1^{-}$ is nilpotent of degree $r$ in $V_B$,  
and $m$ be an integer with $m \ge r$. Let 
us take a positive integer $p$ satisfying $p \le m$. 
We have 
\be 
({ x}_{1}^{-})^{m-p} { x}_{k_1}^{-} \cdots { x}_{k_p}^{-} \Psi   
= A_{k_1, \cdots, k_p}^{(r)}  \, ({ x}_1^{-})^{m} \Psi  \, ,  
\label{r-p}
\ee 
for any set of positive integers $k_1, \ldots, k_p$. 
\label{lem:inductive}
\end{lemma} 
\begin{proof}  
We prove (\ref{r-p}) by induction on $p$ by making use of 
eq. (\ref{eq:null-rel}). 
\end{proof}

\begin{lemma} 
The following  recursive formulas on $n$ hold for $n > 0$:   
\begin{itemize}  
\item[$({A}_n)$:] \quad  
$({ x}_{0}^{+})^{(n-1)} ({ x}_{1}^{-})^{(n)}   
= \sum_{j=1}^{n} (-1)^{j-1} { x}_{j}^{-} 
({ x}_{0}^{+})^{(n-j)} 
({ x}_{1}^{-})^{(n-j)}$ {\rm mod} $U(B) B_{+}$. 
\item[$({B}_n)$:] \quad 
$ n \, ({ x}_{0}^{+})^{(n)} ({ x}_{1}^{-})^{(n)} 
= \sum_{j=1}^{n} (-1)^{j-1} { h}_{j} 
 ({ x}_{0}^{+})^{(n-j)} ({ x}_{1}^{-})^{(n-j)}$ 
{\rm  mod} $U(B) B_{+}$ .  
\item[$({C}_n)$:] \quad 
${[} { h}_1,  ({ x}_{0}^{+})^{(m)} 
({ x}_{1}^{-})^{(m)} {]} = 0 $ \, 
{\rm mod} $U(B) B_{+}$ \, \, for $m \le n$. 
\end{itemize} 
\label{lem:ABC-0}
\end{lemma} 

Making use of $(B_{n})$ of lemma \ref{lem:ABC-0} inductively,   
we show that $\Psi$ is a simultaneous eigenvector of operators 
 $({ x}_{0}^{+})^{(n)}({ x}_{1}^{-})^{(n)}$ for $n>0$.  
 For a given positive integer $k$, 
 we denote by  $\lambda_k$  the eigenvalue:   
 $({ x}_{0}^{+})^{(k)}({ x}_{1}^{-})^{(k)} \Psi =  \lambda_k \Psi$.   
\begin{lemma} 
If ${ x}_1^{-}$ is nilpotent of degree $r$ in $V_B$, 
we have  
\be 
{ x}_{r+1}^{-} \, \Psi = \sum_{j=1}^{r} (-1)^{r-j} 
\lambda_{r+1-j} { x}^{-}_{j} \, \Psi  \, . 
\label{red-rel}
\ee
Moreover, it leads to the following: 
\be 
{ x}_{r+1+p}^{-} \, \Psi = \sum_{j=1}^{r} (-1)^{r-j} 
\lambda_{r+1+p-j} { x}^{-}_{j+p} \, \Psi  \, , 
\quad {\rm for } \, \, p \in {\bf Z}_{\ge 0}. 
\label{red-rel-p}
\ee
 \label{lem:red-rel}
\end{lemma} 
\begin{proof} 
Relation (\ref{red-rel}) is derived from 
$(A_{r+1})$ of lemma \ref{lem:ABC-0}.   
Making use of    
${ x}_{r+1+n}^{-}  = (-2)^{-1} [{ h}_{n},  { x}_{r+1}^{-}] $ 
and (\ref{red-rel}), we derive   (\ref{red-rel-p}). 
\end{proof} 

\begin{proposition}
Suppose that ${ x }_1^{-}$ is nilpotent of degree $r$ in $V_B$.     
In the sector of ${ h}_0 = d_0 - 2n$ with $0 \le n \le r$,   
every vector is expressed as a sum of monomial vectors 
 ${ x}_{k_1}^{-} \cdots { x}_{k_n}^{-} \Psi$ 
for integers  $k_1,  k_2, \ldots,  k_n$ satisfying  
$1 \le k_1 \le k_2 \le \cdots \le  k_n \le r$.  
\label{prop:sectoring}
\end{proposition}
\begin{proof}
It is clear from (\ref{red-rel-p}).  
\end{proof} 


\setcounter{section}{2} 
 \setcounter{equation}{0} 
 \renewcommand{\theequation}{3.\arabic{equation}}
\section{Generalized Drinfeld Polynomials $P_{\Psi}(u)$ for $V_B$} 

%
\begin{definition}
Suppose that ${ x}_1^{-}$ is nilpotent of degree $r$ in $V_B$. 
We define a polynomial $P_{\Psi}(u)$ by 
\be
P_{\Psi}(u) = \sum_{k=0}^{r} \lambda_k (-u)^k  \, . 
\ee
\label{df:PV(u)}
\end{definition} 

\begin{definition}
If polynomial $P_{\Psi}(u)$ of $V_B$ 
is factorized as  
\be 
P_{\Psi}(u)= \prod_{k=1}^{s} (1 - a_k u)^{m_k} \, ,   
\label{eq:factorization}
\ee
where $a_1, a_2, \ldots, a_s$ are distinct, and their  
 multiplicities are given by  $m_1, m_2, \ldots, m_s$, respectively,    
then we call  $a_j$ the {\it evaluation parameters} of 
highest weight vector $\Psi$. 
 We denote by ${\bm a}$ the set of 
$s$ parameters, $a_1, a_2, \ldots, a_s$. 
\label{df:eval-pa}
\end{definition}

We note that $r$ is given by the sum: $r=m_1 + \cdots + m_s$. 
Let us define parameters ${\hat a}_i$ 
for $i=1, 2, \ldots, r$, as follows:  
\be 
{\hat a}_i = a_k \quad {\rm if } \, \, m_1+ m_2 + \cdots + m_{k-1} < i \le  
m_1+ \cdots + m_{k-1} + m_{k} \, . 
\label{eq:hat-a}
\ee
Then, the set 
${\bm {\hat a}}= \{  {\hat a}_j \, | j =1, 2, \ldots, r \}$ corresponds to the 
set of evaluation parameters $a_j$ with multiplicities $m_j$ for 
 $j=1, 2, \ldots, r$.

\section{Generators with parameters }

\subsection{Loop algebra generators with parameters}

Let $A$ be a set of parameters such as 
$\{\alpha_1, \alpha_2, \ldots, \alpha_m \}$. 
We define generators with $m$ parameters $x_m^{\pm}(A)$ and 
$h_m(A)$ as follows \cite{RAQIS05}: 
\begin{eqnarray} 
x_{m}^{\pm}(A) & = & 
\sum_{k=0}^{m} (-1)^k x_{m-k}^{\pm} 
\sum_{\{ i_1, \ldots, i_k \} \subset \{1, \ldots, m \}}  
\alpha_{i_1} \alpha_{i_2} \cdots \alpha_{i_k} \, , 
\non \\
h_{m}(A) & = & 
\sum_{k=0}^{m} (-1)^k h_{m-k} 
\sum_{\{ i_1, \ldots, i_k \} \subset \{1, \ldots, m \}}  
\alpha_{i_1} \alpha_{i_2} \cdots \alpha_{i_k} \, . 
\end{eqnarray}
In terms of generators with parameters 
we generalize the defining relations of the $sl_2$ loop algebra. 
Let $A$ and $B$ are arbitrary sets of $m$ and $n$ parameters, 
respectively.  The operators with parameters satisfy the following: 
\begin{equation} 
[x_m^{+}(A), x_n^{-}(B)] = h_{m+n}(A \cup B) \, , \quad 
[h_{m}(A), x^{\pm}_n(B)] = \pm 2  x_{m+n}^{\pm}(A \cup B) \, . 
\label{eq:dfr-AB}
\end{equation}
By using  the relations (\ref{eq:dfr-AB}), 
it is straightforward to show the following: 
\begin{eqnarray} 
 [ x_{\ell}^{+}(A), (x_m^{-}(B))^{(n)} ] & = & (x_m^{-}(B) )^{(n-1)} 
 h_{\ell+m}(A \cup B) \nonumber \\
 & &  - x_{\ell+2m}^{-}(A \cup B \cup B) 
 (x_{m}^{-}(B))^{(n-2)} \, , \non \\ 
 {[} h_{\ell}(A), (x_m^{\pm}(B))^{(n)} {]} & = & \pm 2 (x_m^{\pm}(B))^{(n-1)} 
 x_{\ell+m}^{\pm}(A \cup B) \, . \label{eq:AB}
\end{eqnarray}
Here the symbol $(X)^{(n)}$ denotes the $n$th power of operator $X$ 
divided by the $n$ factorial, i.e. $(X)^{(n)} = X/n!$ .

Let the symbol ${\bm \alpha}$ denote 
a set of $m$ parameters, $\al_j$ for $j=1, 2, \ldots, m$.  
We denote by $A_{j}$ the set of all the parameters  
except for $\al_j$, i.e.  
$A_{j}= {\bm \alpha} \setminus \{\al_j  \}= 
\{\al_1, \ldots, \al_{j-1}, \al_{j+1}, \ldots, \al_m \}$.   
We introduce the following symbol: 
\begin{equation} 
\rho_j^{\pm}({\bm \alpha}) = x_{m-1}^{\pm}(A_j) \, \quad 
{\rm for } \quad j = 1, 2, \ldots, m . 
\end{equation} 

Here we note the following:  
\begin{lemma}
If $x_{n}^{-}(A)\Omega=0$ for some set of $n$ parameters, $A$, 
then we have $x_{n+m}^{-}(A\cup B)\Omega=0$ for any 
set of $m$ parameters, $B$. 
\label{lem:vanish}
\end{lemma}

Hereafter, we denote by $a_j^{\otimes m}$ 
the set of parameter $a_j$ with multiplicity $m$, 
i.e. $a_j^{\otimes m}= \{ a_j, a_j, \ldots, a_j \}$.  
Moreover, in the case of $m=1$, we write $x_1^{\pm}(a_j^{\otimes 1})$ 
simply as $x_1^{\pm}(a_j)$. 
  
\subsection{Borel subalgebra generators with parameters}

In the case of the Borel subalgebra $U(B)$, 
we do not have generator $x_0^{-}$ in $U(B)$. 
In order to introduce generators with parameters for $U(B)$, 
we thus need some trick.     

For a given set of $m$ parameters, $\al_j$ for $j=1, 2, \ldots, m$, 
we introduce the extended set of parameters as follows: 
\be 
{\bm \alpha}^{(n)} = {\bm \alpha} \cup \{ 0^{\otimes n} \} \, . 
\ee
Here we recall that $a^{\otimes n}$ denotes the set of $a$ 
with multiplicity $n$.  
We also introduce the following symbols: 
\begin{equation} 
\rho_j^{\pm}({\bm \alpha}^{(1)}) = x_{m}^{\pm}(A_j^{(1)}) \, \quad 
{\rm for } \quad j = 1, 2, \ldots, m . 
\end{equation} 
It is easy to show    
\be 
\sum_{j=1}^{n} {\frac {\rho_j^{\pm}({\bm \alpha}^{(1)})} 
{\prod_{k=1; k \ne j}^{m} \al_{kj}} } 
= x_{m+1-n}^{\pm}(\{ \al_{n+1}, \ldots, \al_{m} \}\cup \{0 \})  
\quad (1 \le n \le m)  \, . 
\ee     
It follows inductively on $n$ that $x_{k}^{-}$  for $1 \le k \le m$  
are expressed in terms of linear combinations of 
$\rho_j^{-}({\bm \alpha}^{(1)})$ with $1 \le j \le m$.

The reduction relation (\ref{red-rel})  
is expressed as $x_{r+1}^{-} ({\bm {\hat a}}^{(1)}) \Psi = 0.$ 
However, if we have  
\be 
x_{s+1}^{-} ({\bm a}^{(1)}) \Psi = 0 \, , 
\label{eq:red-s}
\ee 
making use of (\ref{eq:red-s}), we can express  
monomial vector $x_{j_1}^{-} x_{j_2}^{-} \cdots x_{j_n}^{-} \Psi$ of 
any set of positive integers, $j_1, \ldots, j_n$,  
as a linear combination of 
$ \rho_{k_1}^{-}({\bm a}^{(1)}) \rho_{k_2}^{-}({\bm a}^{(1)}) \cdots
 \rho_{k_n}^{-}({\bm a}^{(1)}) \Psi$ 
over some sets of integers with  
$1 \le k_1, \ldots, k_n \le r$.

\section{Highest weight representations}

\subsection{The case of distinct evaluation parameters}

Let us discuss the case where all the evaluation parameters 
$a_j$ have multiplicity 1, i.e. $m_j=1$ for $j=1, \ldots, s$.  
We call it  the case of distinct evaluation parameters. 
Here we note that $s=r$. We therefore have 
\be 
x_{s+1}^{-} ({\bm a}^{(1)}) \Psi = 0 \, . \label{eq:red-distinct}
\ee
 
\begin{lemma} If all evaluation parameters ${\hat a}_j$ 
are distinct ($m_j=1$ for all $j$), we have 
\be 
\left(\rho_j^{-}({\bm a}^{(1)})\right)^2 \, \Psi = 0 \, . 
\label{eq:vanish} 
\ee
\label{lem:square0}
\end{lemma}
\begin{proof} 
 First, we show 
 \be 
 x_0^{+} \,  (\rho_j^{-}({\bm a}^{(1)}))^2 \, \Psi = 0 \, . \label{eq:x0}
 \ee
 From eq. (\ref{eq:AB}) we have 
$$ 
x_0^{+} \, (\rho_j^{-}({\bm a}^{(1)}))^{(2)} \Psi 
= x_{s}^{-}(A_j^{(1)}) h_{s}(A_j^{(1)}) \Psi 
- x_{2s}^{-} (A_j^{(1)} \cup A_j^{(1)}) \Psi \, . 
$$
We set $a_0 = 0$. In terms of $a_{kj}=a_k - a_j$, we have 
$$
h_{s}(A_j^{(1)}) \Psi = \prod_{k=0; k \ne j}^{s} a_{jk} \, \Psi \, ,  
$$ 
and using eq. (\ref{eq:red-distinct}) and lemma \ref{lem:vanish} 
we have 
$$
x_{2s}^{-} (A_j^{(1)} \cup A_j^{(1)} ) \Psi = 
a_{j0} \, \prod_{k=1 ;\ne j}^{s} a_{jk} \, 
x_{s}^{-}(A_j^{(1)}) \Psi \, . 
$$ 
We thus obtain eq. (\ref{eq:x0}).  
Secondly, we apply $(x_0^{+})^{(r-1)}(x_1^{-}(a_j))^{(r-1)}$ to 
$(\rho_j^{-}({\bm a}^{(1)}))^2 \Psi$. The product is given by zero 
since it is out of the sectors of $V_{\Psi}$ due to the fact that 
$(r-1)+2 > r$ and proposition \ref{prop:nilpotent}:    
$$ 
(x_0^{+})^{(r-1)}(x_1^{-}(a_j))^{(r-1)} \, 
(\rho_j^{-}({\bm a}^{(1)}))^2 \, \Psi= 0 \, . 
$$ 
We then show that the left-hand-side is given by 
$$
 \rho_j^{-}({\bm a}^{(1)})^2 \, 
 (x_0^{+})^{(r-1)}(x_1^{-}(a_j))^{(r-1)} \, \Psi 
 = \prod_{k=1; k \ne j}^{r} a_{kj} \, \times \,
  (\rho_j^{-}({\bm a}^{(1)}))^2 \, \Psi \, . 
$$
Here, through induction on $n$ 
and using $B_n$ of lemma (\ref{lem:ABC-0}), 
we show 
$$
[(x_0^{+})^{(n)}(x_1^{-}(a_j))^{(n)}, 
\rho_j^{-}({\bm a}^{(1)})^2 ] \Psi =0 \,  
\quad  (n \le r-1) .  
$$
Since $a_{kj} \ne 0$ for $k \ne j$, 
 we obtain eq. (\ref{eq:vanish}).  
\end{proof}

\begin{lemma} 
Let $x_1^{-}$ be nilpotent of degree $r$ in $V_B$.  
In the sector of $h_0=d_0-2n$ for an integer $n$ with $0 \le n \le r$,  
every vector $v_n$ is written as   
\be 
v_n = \sum_{1 \le j_1 < \cdots < j_n \le s} 
C_{j_1, \cdots, j_n} \, \prod_{t=1}^{n} 
\rho_{j_t}^{-}({\bm a}^{(1)}) \,  \Psi \, . 
\label{eq:vnC}
\ee
Suppose that $\lambda_r \ne 0$.  Then, if $v_n$ is zero, 
all the coefficients $C_{j_1, \cdots, j_n}$ in (\ref{eq:vnC}) 
are given by zero.  
\label{lem:vnC} 
\end{lemma}
\begin{proof} 
In terms of $\rho_j^{-}({\bm a}^{(1)})$, any vector in the sector 
is expressed as a linear combination of  
$\rho_{j_1}^{-}({\bm a}^{(1)})  \cdots 
\rho_{j_n}^{-}({\bm a}^{(1)}) \,  \Psi$.   From lemma 
\ref{lem:square0} we may assume $1 \le j_1 < \cdots < j_n \le s$. 
For a set of integers with $1 \le i_1, \ldots, i_n \le s$,   
multiplying  both sides of eq. (\ref{eq:vnC}) 
with $\rho_{i_1}^{+}({\bm a}^{(1)})  \cdots 
\rho_{i_n}^{+}({\bm a}^{(1)})$, we have   
$$
\rho_{i_1}^{+}({\bm a}^{(1)})  \cdots 
\rho_{i_n}^{+}({\bm a}^{(1)}) v_n = C_{i_1, \cdots, i_n} \,  
\prod_{t=1}^{n} \prod_{k=0; k \ne i_t}^{s} a_{i_t k}^2  
\, \times \, \Psi  
$$
Therefore, if $v_n=0$,  all the coefficients 
$C_{j_1, \cdots, j_n}$ are given by zero. 
\end{proof} 

 From lemmas \ref{lem:square0}, \ref{lem:vnC} and 
proposition \ref{prop:nilpotent} 
we have the following: 
\begin{prop} If  evaluation parameters ${\hat a}_j$ 
of $\Psi$ are distinct, the set of vectors 
$\prod_{t=1}^{n} \rho_{j_t}^{-}({\bm a}^{(1)}) \, \Psi$ for 
$1 \le j_1 < \cdots < j_n \le s$ gives a basis 
of the sector of $h_0=d_0 - 2n$ in $V_B$.    
\label{prop:basis}
\end{prop}

\begin{theorem}
Let $V_B$ denotes the finite-dimensional representation of $U(B)$ 
 generated by a highest weight vector $\Psi$. 
If  $x_1^{-}$ is nilpotent of degree $r$ in $V_B$ and 
$\Psi$ has distinct and nonzero evaluation parameters $a_1, \ldots, a_r$,  
then $V_B$ is irreducible. 
\label{th:distinct}
\end{theorem}
\begin{proof} 
We show that every nonzero vector of $V_B$ has such an element 
of the loop algebra 
that maps it to $\Psi$. Suppose that there is a nonzero vector $v_n$ 
in the sector of $h_0=d_0-2n$ that has no such element. 
Then, we have  
 \be 
 x_{k_1}^{+} \cdots x_{k_n}^{+} \, v_n =0
\label{eq:vn0}
 \ee 
for all monomial elements $x_{k_1}^{+}\cdots x_{k_n}^{+}$.  
Here  $v_n$ is expressed in terms of the basis vectors 
$\rho_{j_1}^{-}({\bm a}^{(1)}) \cdots \rho_{j_n}^{-}({\bm a}^{(1)}) 
\Psi$ with coefficients $C_{j_1, \ldots, j_n}$ and 
$1 \le j_1 < \cdots < j_n \le s$, 
as in  (\ref{eq:vnC}). Then,  
by the same argument as in lemma \ref{lem:vnC} 
 we show that all the coefficients $C_{j_1, \ldots, j_n}$ vanish. 
However, this contradicts with the assumption that 
$v_n$ is nonzero. It therefore follows that $v_n$ has such an element 
that maps it to $\Psi$. We thus obtain the theorem.  
\end{proof}

\subsection{The case of degenerate evaluation parameters}

Let us discuss a general criteria for a finite-dimensional 
highest weight representation to be irreducible. 
\begin{theorem} 
Recall that $V_B$ is a finite-dimensional representation 
of the Borel subalgebra $U(B)$  
generated by a highest weight vector $\Psi$ 
that has evaluation parameters $a_j$ with multiplicities 
$m_j$ for $j=1, 2, \ldots, s$.  
Suppose that $x_1^{-}$ is nilpotent of degree $r$ and the 
evaluation parameters are nonzero, i.e. 
$a_1 a_2 \cdots a_s \ne 0$. We also recall that 
 ${\bm a}$ denotes the set of evaluation parameters:   
${\bm a}=\{a_1, a_2, \ldots, a_s  \}$. Then,  
$V_{B}$  is irreducible if and only if 
$x_{s+1}^{-}({\bm a}^{(1)}) \Psi = 0$.   
\label{th:degenerate}
\end{theorem}
We prove it by generalizing 
the proof of theorem \ref{th:distinct} (cf. Ref. \cite{RAQIS05}).

Theorem \ref{th:degenerate} 
 plays an important role when we discuss 
the spectral degeneracy of the twisted XXZ spin chain at roots of unity 
associated with the Borel subalgebra $U(B)$ of 
the $sl_2$ loop algebra. Here the spin chain 
satisfies the twisted boundary conditions. 
We show in some sectors  that a regular Bethe ansatz eigenvector 
$|R; \Phi \ra$ 
is a highest weight vector of the Borel subalgebra $U(B)$ 
for some twist angle $\Phi$ 
\cite{Regular,Deguchi-Kudo}.  
It is nontrivial whether the 
highest weight representation $V_B$ generated by $|R; \Phi \ra$ 
is irreducible or not. 
Suppose that $x_1^{-}$ is nilpotent of degree $r$ in $V_B$,  
$|R; \Phi \ra$ has nonzero evaluation parameters $a_j$ with multiplicities 
$m_j$ for $j=1, 2, \ldots, s$, where $m_1 + \cdots + m_s = r$, 
and we have 
the following relation:  
\be 
x_{s+1}^{-}({\bm a}^{(1)}) \, |R; \Phi \ra = 0 \, ,  
\label{eq:criteria}
\ee 
where ${\bm a}$ denotes 
the set of evaluation parameters $a_1, a_2, \ldots, a_s$.  
Then, it follows from theorem \ref{th:degenerate} that 
$V_B$ is irreducible, and the degenerate multiplicity 
of $|R; \Phi \ra$ is given by 
$(m_1+1)(m_2 +1)  \cdots (m_s + 1)$.

\section*{Acknowledgments}
The author would like to thank 
the organizers for their kind hospitality and invitation to 
the 23rd International Conference 
of Differential Geometric Methods 
in Theoretical Physics, August 20-26, 2005, 
Nankai Institute of Mathematics, Tianjin, China. 
He is also like to thank Dr. A. Nishino for his interest in the research. 
This work is partially supported by 
Grant-in-Aid for Scientific Research (C) No. 17540351.


\end{document}